\documentstyle[referee]{laa}

\begin{document}
\thesaurus{07  %A&A Section: Solar system
	   (07.09.1; % Interplanetary medium
	    07.13.1;  % Meteoroids
	   )}

\title{On physics of the Poynting-Robertson effect}
\author{J.~Kla\v{c}ka}
\institute{Institute of Astronomy,
   Faculty for Mathematics and Physics, Comenius University \\
   Mlynsk\'{a} dolina, 842~48 Bratislava, Slovak Republic}
\date{}
\maketitle

\begin{abstract}
Detailed discussion of university textbook statements (Harwit 1988)
concerning the Poynting-Robertson effect (P-R effect) is presented. Discussion
is concentrated on better physical understanding of the P-R effect.
References to complete correct equation of motion for real dust particle
are also presented.

\end{abstract}

\section{Introduction}
Orbital evolution due to the interaction between cosmic dust particle and
electromagnetic radiation is -- as most astronomers think -- well-known as the
Poynting-Robertson effect (P-R effect).

Srikanth (1999) offers three physical viewpoints on the corresponding
statements presented in astronomical literature which is the most referenced on
the P-R effect. However, more detailed physics of the P-R effect was presented
by Kla\v{c}ka (1992). The statements of Srikanth (1999) were discussed
by Kla\v{c}ka (2000a).

Astronomers will not understand the real physics of the P-R effect
if incorrect statements are presented in frequently used university textbooks
and scientific papers.
The aim of this paper is to discuss the statements presented in
Harwit (1988, pp. 176-177).

\section{P-R effect and Harwit's discussion}
Harwit (1988) discusses the derivation of the P-R drag in two ways.
The first access concentrates to the situation as seen from the frame
of reference of the Sun. The second access concentrates to the situation as
seen from the frame of reference of the dust particle -- grain.
It is supposed that particle absorbs sunlight and re-emits this energy
isotropically in its own rest frame.

\subsection{Reference frame of the Sun}
Harwit (1988) formulates heuristic three point instruction for obtaining loss
of orbital angular momentum. The most significant incorrect statement is that
"a re-emitted photon carries off angular momentum" in the way that produces
P-R drag. Another heuristic statement concerns the fact that the loss of
angular momentum is proportional to the velocity of the grain -- why?
Eqs. (5-45) and (5-46) are incorrect.

If we want to put the statements into a correct physics, we have to
stress that photons are re-emitted isotropically -- total re-emitted
momentum is zero in the proper frame of reference of the particle.
As a consequence, the particle (grain) losses the momentum proportional to
the velocity of the grain -- consequence of the
Lorentz transformation.

\subsection{Reference frame of the grain}
The access of the Harwit is instructive. The only improvement may
concern the fact that we do not need to consider relativistic aberration
of light if we are considering only terms linear in $v / c$. Moreover,
one should explain the possibility of mixing quantities measured in
two different reference frames (of the sun and of the grain).

\section{P-R effect and Newton's laws of motion}
Harwit (1988) concentrates to explanation of the P-R drag. However,
P-R drag is only a part of the total equation of motion for dust
particle due to its interaction with electromagnetic radiation
(see also Kla\v{c}ka 2000a with respect to the discussion presented by
Srikanth 1999). Thus, complete equation of motion should be derived.

Harwit (1988) uses angular momentum as a relevant physical quantity
-- its one component. However, angular momentum vector cannot
fully describe any general motion. Since the time of Newton we know
that equation of motion is significant and it is described by
time derivative of momentum.

The physical access of Harwit enables to treat the P-R effect only for
near circular orbits. Of course, even this type of orbits is not
correctly -- completely -- described, as it is for the true P-R effect
(see Kla\v{c}ka and Kaufmannov\'{a} 1992; the same results were obtained by
Breiter and Jackson 1998 -- one must be careful, since the
Breiter and Jackson 's result concerning the systematic increase of eccentricity
is not physically correct -- it is caused by linearizing of the P-R effect,
see Kla\v{c}ka 1999).

\section{Physics of the velocity decrease}
Harwit (1988) states that "the grain velocity decreases on just absorbing the
light". This statement is incorrect. We will show it
using generalization of Robertson's (1937) result.

The generalization for the P-R effect is defined by the equation
\begin{equation}\label{1}
\vec{p'_{o}} =  \left ( 1 ~-~ Q'_{PR} \right ) ~ \vec{p'_{i}} ~,
\end{equation}
which corresponds to the case when the total momentum per unit time of the
``outgoing'' radiation $\vec{p'_{o}}$ is proportional to the ``incident''
momentum per unit time $\vec{p'_{i}}$;
primes denote quantities measured in the proper frame of
reference of the particle (see Eq. (122) in Kla\v{c}ka 1992a).
On the basis of Eq. (1) one comes to the following equation
\begin{equation}\label{2}
\frac{d p^{\mu}}{d \tau} = p_{i}^{\mu} ~-~ \left [
		\left ( 1 ~-~ Q'_{PR} \right ) ~p_{i}^{\mu} ~+~
		Q'_{PR} \frac{w ~E_{i}}{c^{2}} ~u^{\mu} \right ] ~,
\end{equation}
where $u^{\mu}$ denotes four-velocity,
$p_{i}^{\mu} = ( E_{i} / c, \vec{p_{i}} )$, $Q'_{PR}$ is pressure coefficient
and the quantity $w ~E_{i}$ is scalar (see Eqs. (30), (133) and (134)
in Kla\v{c}ka 1992).

Let the perfect absorption of light occurs. \\
i) The case of
perfect absorption and no reemission is described by Eq. (2) when the only
first term on the right-hand side is present. Eq. (2) yields then
\begin{eqnarray}\label{3}
\frac{d m}{d \tau} &=& p_{i}^{\nu} ~ u _{\nu} / c^{2} ~,
\nonumber \\
m ~ \frac{d u^{\mu}}{d \tau} &=& p_{i}^{\mu} ~-~ \left (
		p_{i}^{\nu} ~ u _{\nu} / c^{2} \right ) ~ u^{\mu} ~.
\end{eqnarray}
Eq. (3) states that the grain velocity decreases on just absorbing the light
(mass of the particle increases due to the absorption of light). \\
ii) The square bracket terms in Eq. (2)
correspond to reemission in the form
described by Eq. (1). Let the reemission is in the form that $Q'_{PR} =$ 0:
Eq. (2) yields $d u^{\mu} / d \tau =$ 0, then ($d m / d \tau =$ 0).
We see that absorption
is not the cause for decreasing of the grain velocity. \\
We can summarize: the grain
velocity decreases due to the fact that
$Q'_{PR}$ is positive value and that Eq. (1) holds.

As another example, one
may easily calculate interaction of electromagnetic radiation with
spherical particle covered by reflecting mirror (specular reflection):
equation of motion is the same as that for perfectly absorbing spherical
particle with isotropic reemission, i. e., $Q'_{PR} =$ 1 -- the grain
velocity decreases, but no absorption exists.

\section{Conclusion}
We have put into a correct physics the statements presented in Harwit (1988).
As for the correct complete derivation of the P-R effect we refer the reader to
Kla\v{c}ka (1992a, 1992b, 1993a, 1993b), as for other discussions to
Kla\v{c}ka (1993c, 2000a). As for interaction between electromagnetic radiation
and nonspherical particle, we refer to
Kocifaj and Kla\v{c}ka (1999) and to
Kla\v{c}ka (2000b, 2000c, 2000d).

\acknowledgements
The paper was
supported by the Scientific Grant Agency VEGA (grant No. 1/7067/20).

\end{document}